\documentclass{ws-ijmpb}
\def\bfr{{\bf r}}

\begin{document}

\markboth{D. A. W. Hutchinson and P. B. Blakie}
{Phase Transitions in Ultra-Cold Two-Dimensional Bose Gases}

%
\catchline{}{}{}{}{}
%

\title{PHASE TRANSITIONS IN ULTRA-COLD\\ 
TWO-DIMENSIONAL BOSE GASES\  }

\author{D. A. W. HUTCHINSON and P. B. BLAKIE}

\address{Department of Physics, University of Otago, \\
P. O. Box 56, Dunedin,
New Zealand\\
hutch@physics.otago.ac.nz}

\maketitle


\begin{abstract}
We briefly review the theory of Bose-Einstein condensation in the two-dimensional
trapped Bose gas and, in particular the relationship to the theory of the homogeneous
two-dimensional gas and the Berezinskii-Kosterlitz-Thouless phase. We obtain a phase
diagram for the trapped two-dimensional gas, finding a critical temperature above
which the free energy of a state with a pair of vortices of opposite circulation is
lower than that for a vortex-free Bose-Einstein condensed ground state. We identify
three distinct phases which are, in order of increasing temperature, a phase coherent
Bose-Einstein condensate, a vortex pair plasma with fluctuating condensate phase and
a thermal Bose gas. The thermal activation of vortex-antivortex pair formation is
confirmed using finite-temperature classical field simulations.
\end{abstract}


\section{Introduction}	

It is now over a decade since Bose-Einstein condensation (BEC) in a dilute atomic gas
was first realised.\cite{bec} In this time remarkable progress has been made in the control and manipulation of these ultra-cold gases. One possibility offered, through the use of light fields in the form of the optical lattice,\cite{lattice} by this control is to reduce the dimensionality of the system. By increasing the confining potentials in one or two directions it is possible to freeze out these degrees of freedom creating, effectively, two-\cite{gorlitz} or one-dimensional\cite{cristiani} gases. The two-dimensional (2D) gas is of particular interest. In general 2D systems often display unique properties, often remarkably different from those of the comparable bulk system. For example, the 2D electron gas, subjected to a perpendicular magnetic field, displays the rich phenomena of the, essentially single particle, quantum Hall effect and, the even more remarkable, many body fractional quantum Hall effect\cite{girvin} with the physical manifestation of, the uniquely two-dimensional, fractional statistics\cite{wilczek} in the quasiparticle excitations. More specifically, it is well known that the uniform 2D Bose gas does not undergo BEC\cite{7,8} at finite-temperature. The interacting homogeneous 2D gas does still undergo a normal-superfluid phase transition\cite{9,10,11} at finite-temperature however, with an order parameter which is only locally phase coherent - the Berezinskii-Kosterlitz-Thouless (BKT) phase. Experimental evidence for this phase transition has been demonstrated in liquid helium thin film\cite{12}, superconducting Josephson-junction arrays\cite{13} and in spin-polarized atomic hydrogen\cite{14}.

In the trapped 2D Bose gas the density of states is modified sufficiently that, for the ideal gas, condensation into a single state can occur at finite temperature.\cite{bagnato} It has been demonstrated however that long-wavelength phase fluctuations may destroy the global phase coherence of the condensate at a temperature still below the ideal gas critical temperature, $T_0$ for BEC with only a {\it quasi-condensate}\cite{16,17,18,19,20,21,22} present. The connection between such a quasi-condensate and the correlated vortex-pair plasma of the BKT phase proves unclear however and forms the basis of the material we attempt to address in the remainder of this paper.

\section{Free Energy of the Vortex-Antivortex Pair}

In order to determine at what temperature thermal activation of vortex-antivortex pairs becomes thermodynamically favourable, we consider the Helmholtz free energy, $F=E-TS$ of $N$ bosonic particles of mass $m$ spatially confined in 2D and via the harmonic potential $V({\bf r})=m\omega^2_\perp r^2/2$, where $\omega_\perp$ is the radial trap frequency. We assume the existence of a macroscopic ground state condensate wavefunction $\psi(\bfr)$, which also serves to describe the order parameter of the system in the BEC phase. Isolating the energy contribution due to the condensate, $E_0(T)$, the total internal energy of the system may be written as $E(T)= E_0(T)+\tilde{E}(T)$. The required
condensate energy $E_0(T)$ is determined by the functional
\begin{equation}
E_0(T)=\int \left(\frac{\hbar^2}{2m}|\nabla\psi(\bfr)|^2 +V(\bfr)|\psi(\bfr)|^2
+ \frac{g}{2}|\psi(\bfr)|^4\right) {\rm d}\bfr
\label{EGY}
\end{equation}
where $g$ is the constant coupling parameter for the particle interactions, and
$N_0=\int |\psi(\bfr)|^2 {\rm d}\bfr$ denotes the number of particles in the condensate.The internal energies for topologically distinct order
parameter configurations are then calculated computationally.\cite{bkt} The entropy contribution to the free-energy difference is due to the multiplicity of order parameter configurations containing a pair of vortices. The statistical weight $W=2\pi R_{TF}^2/\xi^2$, is obtained by allowing one vortex to reside anywhere within the Thomas-Fermi radius, the partner vortex then having $2 \pi$ available nearest neighbour sites. Assuming the radius of the area occupied by each vortex to be of the order of the healing length, $\xi=\sqrt{\hbar^2/2m\mu}$, we can obtain an expression for the configurational energy difference $\Delta S(T)=k_B\ln{(8\pi\mu^2/\hbar^2\omega_\perp^2)}$. The free energy for each configuration can now be evaluated, yielding a critical temperature for the thermal activation of vortex-antivortex pair creation. For the experimental parameters of the Oxford experiment\cite{chris} we obtain a critical temperature $T_{\rm c} \approx 0.5 T_0$ although the precise value is dependent upon the total number of atoms and decreases monotonically with $N$. Of course taking the thermodynamic limit in this case, which one would need to do if one were to try to claim that this is a phase transition, is slightly complicated by the presence of the trap. 

This indicates that there are three distinct regions for the harmonically trapped, ultra-cold, dilute Bose gas. At low temperatures there is a phase coherent BEC with a coherence length comparable with the system size as shown by Gies and Hutchinson.\cite{gies} At higher temperatures, where the coherence length reduces and off-diagonal order decays algebraically, it becomes thermodynamically favourable to form correlated vortex-antivortex pairs, which we take as a signature of a BKT-like phase. At temperatures approaching the ideal gas critical temperature, $T_0$ the off-diagonal order decays exponentially. At this point the vortex pairs unbind (although this is not present in our free energy model) and superfluidity is destroyed yielding a thermal gas. This is what is traditionally referred to as the BKT transition.

\section{Classical Field Simulations}

To confirm these conclusions finite temperature classical field simulations\cite{matt,blair} were performed. In the classical field approach the system is divided into classical and incoherent regions, determined by the occupation of single particle modes. Highly occupied (classical) modes are described by the projected Gross-Pitaevskii equation
\begin{equation}
i\hbar \frac{\partial \psi(\bfr, t)}{\partial t} = \left[- \frac{\hbar^2}{2m}\nabla^2
+V(\bfr)\right]\psi(\bfr, t) +g {\cal P}\left\{|\psi(\bfr, t)|^2\psi(\bfr, t)\right\}
\end{equation}
where the projector, ${\cal P}$, restricts evolution of the classical field to within its subspace. The low occupation, incoherent modes are described using the semiclassical Hartree-Fock approximation, with the classical and incoherent regions being taken to be in thermal equilibrium with one another. At the lowest temperatures no vortices are present in the system. These first emerge, as the temperature is increased, in the low density regions at the edge of the cloud. At higher temperatures it becomes possible for vortex-antivortex pairs to nucleate closer to the trap centre. This is demonstrated in Fig.(1) which is at a temperature of $0.86 T_0$. 
\begin{figure}[th]
\centerline{\psfig{file=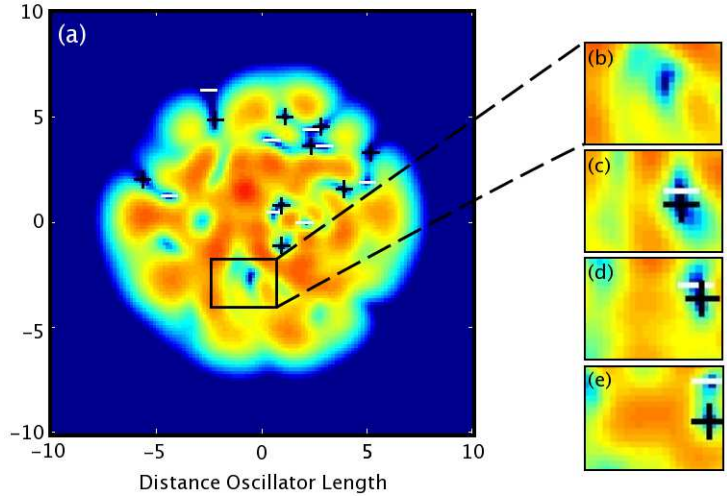,width=10cm}}
\vspace*{8pt}
\caption{Panel (a) shows a single frame of the full simulation with vortices and
antivortices shown by the black crosses and white dashes. The region bounded by the box is blown up into panel (b). Subsequent time frames are shown for this expanded region in panels (c), (d) and (e) clearly showing the vortex-antivortex pair nucleation.}
\end{figure}
Panel (a) shows a single frame of the full simulation with vortices and antivortices shown by the black crosses and white dashes. These are identified through the singularity in the phase as determined from the simulation. The region containing a slight density dip (due to thermal fluctuations) bounded by the box is then blown up into panel (b). Subsequent time frames are then shown for this expanded region in panels (c), (d) and (e). It can clearly be seen how a vortex-antivortex pair forms in the low density fluctuating region, then separates and stabilises. At later time this pair goes on to annihilate again. The temperature at which vortex-antivortex pairs begin to form in the classical field simulations is entirely consistent with the free energy calculations described above and a strong corroboration of the interpretation of the regime with only algebraic off-diagonal order as a BKT-like phase. Further details of the classical field simulations for the 2D gas and comparison with recent experiments\cite{jean} can be found in Ref.~\refcite{tapio}.

\section{Conclusions}

In conclusion we have demonstrated, both using a free energy argument and using classical field simulations that there exist three distinct regions in the phase diagram of the ultra-cold, harmonically confined 2D Bose gas. At very low temperatures compared to the ideal gas condensation temperature $T_0$ the gas forms a phase coherent BEC. At temperatures of order $0.5 T_0$ it becomes thermodynamically favourable for correlated vortex-antivortex pairs to form. The system remains superfluid with algebraic off-diagonal order. This corresponds to a BKT-like superfluid phase. At higher temperatures, approaching $T_0$, the vortex pairs unbind, the off-diagonal density matrix decays exponentially and the superfluid phase is destroyed yielding a thermal gas. This upper transition is what corresponds to the BKT transition in the uniform system.

\section*{Acknowledgements}

We would like to thank Tapio Simula for his contributions to this work. We are also
indebted to the Marsden Fund and to the University of Otago for financial support.

\end{document}